\documentclass[twocolumn,showpacs,preprintnumbers,amsmath,amssymb]{revtex4}

\usepackage{graphicx}
\usepackage{comment}

\begin{document}

\title{Thermalization away from Integrability and the Role of Operator Off-Diagonal Elements}

\author{N. P. Konstantinidis}

\affiliation{Max-Planck-Institut f\"ur Physik komplexer Systeme, 01187 Dresden, Germany}

\date{\today}

\begin{abstract}
We investigate the rate of thermalization of local operators in the one-dimensional anisotropic antiferromagnetic Heisenberg model with next-nearest neighbor interactions that break integrability. This is done by calculating the scaling of the difference of the diagonal and canonical thermal ensemble values as function of system size, and by directly calculating the time evolution of the expectation values of the operators with the Chebyshev polynomial expansion. Spatial and spin symmetry is exploited and the Hamiltonian is divided in subsectors according to their symmetry. The rate of thermalization depends on the proximity to the integrable limit. When integrability is weakly broken thermalization is slow, and becomes faster the stronger the next-nearest neighbor interaction is. Three different regimes for the rate of thermalization with respect to the strength of the integrability breaking parameter are identified. These are shown to be directly connected with the relative strength of the low and higher energy difference off-diagonal operator matrix elements in the symmetry eigenbasis of the Hamiltonian.
Close to the integrable limit the off-diagonal matrix elements peak at higher energies and high frequency fluctuations are important and slow down thermalization. Away from the integrable limit a strong low energy peak gradually develops that takes over the higher frequency fluctuations and leads to quicker thermalization.

\end{abstract}

\pacs{05.30.-d,05.30.Ch,05.70.Ln,75.10.Pq}

\maketitle

\section{Introduction}
\label{sec:Introduction}
The non-equilibrium dynamics of isolated quantum systems has recently been an area of extensive investigation \cite{Srednicki94,Deutsch91,Rigol08,Cazalilla12,Rigol07,Calabrese11,Reimann08,Fagotti13,Pozsgay13}.
A part of an isolated system feels the rest of the system acting as a bath, and is in general in a mixed state. Its non-unitary time evolution can lead in the long-time limit to a thermalized state where fluctuations of operator expectation values are weak, and this state can be described by an appropriate statistical ensemble depending only on a small number of parameters.

The conserved quantities of the Hamiltonian are very important for thermalization. Classical systems where only the energy is conserved are ergodic and thermalize, and the long time average of local observables can be replaced by a time-independent canonical thermal ensemble average. This describes expectation values in the long-time limit and is equivalent to the microcanonical ensemble in the thermodynamic limit (TDL). In contrast, integrable classical systems are non-ergodic. The number of conservation laws scales linearly with system size and they confine the motion in phase space to invariant tori satisfying the conservation laws. These conserved quantities have to be taken into account in the generalized Gibbs ensemble that describes expectation values in the long-time limit. When a weak integrabilty breaking term is applied the Kolmogorov-Arnold-Moser theorem states that some of the invariant tori are deformed and survive \cite{Arnold97}.

A formal prescription for the onset of thermalization when going away from the integrable limit lacks for quantum systems. The prevailing paradigm is that similarly to the classical case non-integrable systems thermalize and their long-time limit is described by a canonical Gibbs ensemble, while the long-time limit of integrable systems is described by a generalized Gibbs ensemble taking into account all conservation laws \cite{Rigol08,Rigol07,Fagotti13,Pozsgay13}.
The rate of thermalization is expected to increase with the distance from the integrable limit. Questions with respect to thermalization are more challenging to answer in quantum mechanics from the analytic as well as the numerical point of view, as with present day computational means it is not possible to calculate the evolution of observables for very long times. Quite recently it has been shown that for weak integrability breaking local operator expectation values go through a prethermalized regime where they first relax to a non-thermal quasi-stationary value before reaching the canonical thermal ensemble value \cite{Kollar11,Essler13}. In general, non-equilibrium studies of strongly correlated models are confined to a relatively small region of parameter space, and very detailed results on the rate of thermalization as function of integrability breaking parameters are not available.

The long-time limit of an operator expectation value is given by the operator's diagonal matrix elements in the energy eigenbasis, which define the diagonal ensemble \cite{Srednicki96,Srednicki99}. Thermalization requires the difference between the diagonal and the canonical Gibbs ensemble to vanish in the TDL. The strength of the fluctuations around the long-time limit is controlled by the off-diagonal terms of the operator matrix, which determine if fluctuations are small enough in order to achieve thermalization, and also which time scales are important for relaxation. Results on this topic have been presented in the literature \cite{Rossini10,Santos10}, and
quite recently the statistical properties of the off-diagonal matrix elements have been studied for two models with respect to their proximity to integrability \cite{Beugeling15}. However, these authors intentionally break spatial symmetry and consequently their results are neither representative of typical connectivities of strongly correlated models nor of individual symmetry sectors that oftentimes solely include the time-evolved wavefunction.

In this paper we investigate the rate of thermalization when integrability breaking is controlled by a parameter. We show how thermalization of local operators, which relates to the time dependence of the fluctuations of the operator expectation values, is directly connected to the strength of the operator off-diagonal matrix elements. We do so within the framework of the anisotropic antiferromagnetic Heisenberg model (AAHM) on chains with $N$ spins of magnitude $s=1/2$ and periodic boundary conditions. The model is taken to have nearest neighbor interactions, and weaker next-nearest neighbor interactions that break integrability in a controlled manner by tuning their strength. We consider the typical in non-equilibrium dynamics case of an interacting quench and take the spatial and spin symmetries of the Hamiltonian into account.
The initial state is the ground state of the AAHM for specific anisotropy, chosen so that the post-quench effective temperature is approximately the same for all values of the integrability breaking parameter of the model.
The Hamiltonian is diagonalized according to its full symmetry group before and after the quench \cite{NPK05,NPK07,NPK09}, and the difference of the diagonal and thermal ensemble values for local operators is calculated as function of system size to establish if it tends to zero at the TDL, as required by thermalization. In addition, the time evolution of the operators is calculated with the Chebyshev polynomial expansion \cite{Zhang07,Zhang08} for $N$ significantly larger than the system maximum that can be exactly diagonalized and for long enough times that show how close the system is to thermalization. The results show how the rate of thermalization increases with the integrability breaking parameter. We also calculate the expectation values of the off-diagonal elements of the operators in question in the symmetry eigenbasis of the Hamiltonian, and more specifically within the symmetry subsector that exclusively includes the wavefunction after the quench. Their distribution as function of the energy difference of the eigenstates is shown to be in direct relation with the rate of thermalization. The observables considered are the spin correlation functions $s_i^x s_j^x + s_i^y s_j^y$ and $s_i^z s_j^z$ that are a maximum of two sites apart $|j-i|=1, 2$ along the closed chain.

We find that for weak integrability breaking the difference between the diagonal and canonical thermal ensemble values decreases sublinearly with $\frac{1}{N}$, indicating slow thermalization. Here it is assumed that the equilibrium value of an operator can be found by first calculating the long time average and then taking the TDL \cite{Sirker14}, which is expected to be adequate for large interacting systems. The corresponding time-dependent expectation values of the operators in question do not differ significantly from the thermal expectation values, but these differences do not decrease significantly for the highest available times which further corroborates a relative slow convergence to the thermal ensemble for weak integrability breaking. When the integrability breaking term gets stronger, roughly one tenth of the nearest neighbor interaction, thermalization becomes faster with the scaling of the difference between the two ensembles very close to linear. Furthermore, the time evolution of the operator expectation values approaches the thermal ensemble with a faster pace. For even stronger integrability breaking the scaling of the difference of the two ensembles becomes superlinear and the time dependent operators approach the thermal ensemble values very closely. We conclude that there are three different regimes with respect to the rate of thermalization when integrability is broken, which are in direct correlation with the strength of the integrability breaking parameter. This conclusion is supported by two different calculations that are in very close agreement with each other.

The rate of thermalization is determined by the strength of the operator off-diagonal elements in the energy eigenbasis, which determine the magnitude of temporal fluctuations. When operator matrix elements between states that differ significantly in energy strongly prevail over their counterparts with small energy differences, high frequency fluctuations are strong and thermalization is slow. The reason is that it takes a considerable time for these fast fluctuations to cancel each other out and be overtaken by the weak low frequency fluctuations that correspond to matrix elements between eigenstates with small energy difference. In contrast, if matrix elements between states not differing much in energy are also of considerable strength high frequency fluctuations diminish faster, leading to quicker thermalization. To explain the thermalization scenario emerging from the comparison between the difference of the diagonal and thermal ensembles and the time evolution of operator expectation values we calculate the distribution of the corresponding operator off-diagonal matrix elements as function of energy difference in the symmetrized energy eigenbasis. In the integrable limit the Hamiltonian subsector where time evolution takes place is characterized by the required spatial, total spin along the $z$ axis $S^z$ and spin inversion symmetries (SU(2) symmetry is in general broken for the AAHM) of the Hamiltonian, however there are more conserved quantities \cite{Grabowski95} which are not taken into account and consequently do not characterize the states in the subsector of the time evolved wavefunction. A Hamiltonian subsector where all symmetries have been taken into account is characterized by level repulsion and has an energy difference between any pair of states which is in general not close to zero, provided there are no Hamiltonian parameters much smaller than the others. On the other hand, in the integrable limit where many of the conservation laws are not taken into account in the block diagonalization of the Hamiltonian it is expected that there will be levels very close in energy, as they would belong to different subsectors of the Hamiltonian if the extra conservation laws were taken into account. Because of that the Hamiltonian matrix elements between such levels should be close to zero. The nearest neighbor correlation functions $s_i^x s_{i+1}^x + s_i^y s_{i+1}^y$ and $s_i^z s_{i+1}^z$ are translationally invariant in singly degenerate subsectors and qualitatively similar to the nearest neighbor energy, consequently their off-diagonal matrix elements between eigenstates very close in energy should also be close to zero in the integrable limit. Hence the off-diagonal elements of these operators have a peak for an energy difference away from zero.

Once the integrability breaking term is switched on the additional conservation laws not taken into account in the integrable case are not fulfilled anymore, consequently states with small energy differences will start mixing more strongly and the correlation functions will develop significant off-diagonal elements between such states. The strength of these off-diagonal elements is regulated by the magnitude of the integrability breaking term. According to low order perturbation theory the mixing is in fact favored for levels which are close in energy in the integrable limit. This results in a peak for small energy differences that now coexists with the higher energy peak that determines the higher frequency fluctuations.  When integrability is weakly broken the low energy peak is relatively weak in comparison with the higher energy peak. This shows that a long time is required for the high frequency fluctuations to cancel each other out and thermalization to settle in, resulting in a long transient regime before thermalization.
For stronger next-nearest neighbor interaction the low energy peak becomes of the same order of magnitude with the higher energy peak, thus the high frequency fluctuations become unimportant faster and thermalization is quicker. This provides the explanation to the results calculated with exact diagonalization and the Chebyshev polynomial expansion. All in all, the non-equilibrium properties of the AAHM with a next-nearest neighbor integrability breaking term are determined by the strength of this term.

The plan of this paper is as follows: In Sec. \ref{sec:Model} the model and the methods to be used are introduced. In Sec. \ref{sec:timeevolution} general considerations about the time evolution of an operator expectation value after a quench are presented. Sec. \ref{subsec:scalingcorrelations} examines the scaling of the difference of the diagonal and canonical ensemble for an interacting quench and also includes time-dependent operator expectation values. Sec. \ref{subsec:offdiagonalterms} correlates thermalization with the strength of operator off-diagonal terms, and finally Sec. \ref{sec:conclusions} presents the conclusions.

\section{Model and Method}
\label{sec:Model}
The Hamiltonian of the AAHM model on a chain with $N$ spins is
\begin{eqnarray}
H & = & \sum_{i=1}^N ( J \vec{s}_i \cdot \vec{s}_{i+1}  + J' \vec{s}_i \cdot \vec{s}_{i+2} )
\label{eq:Hchain}
\end{eqnarray}
The spin magnitude is taken to be $s=1/2$ and periodic boundary conditions are assumed so that $s_{N+1} \equiv s_1$, $s_{N+2} \equiv s_2$. $N$ is taken to be even. $J$ controls the strength of nearest neighbor interactions, and $J'$ the strength of next-nearest neighbor interactions that break integrability. Both of them are taken positive and thus support antiferromagnetic correlations, with $J'$ introducing frustration. The interaction along the spin $z$ axis is scaled by $\Delta$, $\vec{s}_i \cdot \vec{s}_j = s_i^x s_j^x + s_i^y s_j^y + \Delta s_i^z s_j^z$. $J=1$ from now on, defining the unit of energy, and $\Delta'=\Delta$.

Calculation of the diagonal and thermal ensemble values requires full diagonalization of Hamiltonian (\ref{eq:Hchain}). To this end, the spatial $D_N$ and the spin inversion symmetry (when $S^z=0$) are exploited, which is particularly advantageous \cite{NPK05,NPK07,NPK09}. The time evolution of various operators can be calculated directly with exact diagonalization, but the more economical method of the Chebyshev polynomial expansion increases the maximum $N$ that can be considered \cite{Zhang07,Zhang08}, especially since the Hamiltonian block-diagonalizes according to the irreducible representations of the total symmetry group and the wavefunction is time-evolved separately within each representation. If the initial wavefunction belongs solely to a specific symmetry subsector, then it time evolves only in this subsector and also the diagonal ensemble contains operator expectation values of this subsector only. This is the case here as the initial state in an interacting quench is the ground state of the initial Hamiltonian and therefore belongs to a specific irreducible representation of the full symmetry group of Hamiltonian (\ref{eq:Hchain}).
The initial state is chosen to be the ground state of Hamiltonian (\ref{eq:Hchain}) for $\Delta_0=15$, chosen so that the post-quench effective temperature is nearly the same for all values of $J'$, with the post-quench $\Delta=1.1$.

Hamiltonian (\ref{eq:Hchain}) is equivalent to the one of the sawtooth chain with variable interactions \cite{Nakamura96,Sen96,Blundell04,Lavarelo14}. If $J'=0$ or $\frac{J'}{J} \to \infty$ it corresponds to the unfrustrated AAHM and is integrable via the Bethe Ansatz \cite{Bethe31}. For $\Delta=1$ it is also integrable for the Majumdar-Ghosh point $\frac{J'}{J}=\frac{1}{2}$ \cite{Majumdar69,Majumdar69-1}. For $\Delta=1$ and small $\frac{J'}{J}$ there is quasi-long range order in the ground state which gives way to a gapped phase characterized by a dimerization order parameter at $\frac{J'}{J}$ slightly less than $\frac{1}{4}$ \cite{Eggert96}. Here we focus on the region of relatively small $\frac{J'}{J} \leq \frac{1}{5}$ to investigate the effect of integrability breaking in thermalization for operators of the form $s_i^x s_j^x + s_i^y s_j^y$ and $s_i^z s_j^z$ with $|j-i|=1, 2$. We pick $\Delta=1.1$ for the post-quench Hamiltonian to avoid the mixing of the different $S$ sectors of the spin isotropic SU(2) case when one works in the $S^z$ basis.


\section{Time Evolution after a Quench}
\label{sec:timeevolution}
The initial wavefunction $\vert \Psi(t=0) \rangle = \vert \Psi(0) \rangle$ is chosen to be the ground state before the quench. Its post-quench time evolution is given by
\begin{eqnarray}
\vert \Psi (t) \rangle = e^{-iHt} \vert \Psi(0) \rangle = \sum_n C_n e^{-iE_nt} \vert \Psi_n \rangle
\end{eqnarray}
with $E_n$ and $\vert \Psi_n \rangle$ the eigenvalues and eigenvectors of Hamiltonian (\ref{eq:Hchain}) after the quench. The coefficients $C_n \equiv \langle \Psi_n \vert \Psi(0) \rangle$ are independent of time. The expectation value of operator $\hat{O}$ at time $t$ is
\begin{eqnarray}
<\hat{O}(t)> & = & \sum_n \vert C_n \vert ^2 \langle \Psi_n \vert \hat{O} \vert \Psi_n \rangle + \nonumber \\ & & \sum_{m \neq n} C_n^* C_m e^{-i(E_m-E_n)t} \langle \Psi_n \vert \hat{O} \vert \Psi_m \rangle
\label{eq:diagonal}
\end{eqnarray}
In the time average of Eq. (\ref{eq:diagonal}) the second term averages out to zero and the average is $\bar{O} \equiv \sum_n \vert C_n \vert ^2 \langle \Psi_n \vert \hat{O} \vert \Psi_n \rangle$, defining the diagonal ensemble $\hat{O}_{diag} \equiv \sum_n \langle \Psi_n \vert \hat{O} \vert \Psi_n \rangle \vert \Psi_n \rangle \langle \Psi_n \vert$ \cite{Rigol08}. In the interacting quench considered here time evolution takes place in non-degenerate subsectors, therefore degeneracies play no role in the second term of Eq. (\ref{eq:diagonal}). For thermalization to occur the diagonal ensemble prediction must equal the value of the canonical thermal ensemble in the TDL.
Eq. (\ref{eq:diagonal}) shows that the strength of the fluctuations around the diagonal ensemble value is determined by the off-diagonal terms $\langle \Psi_n \vert \hat{O} \vert \Psi_m \rangle \equiv \hat{O}_{nm}$, $m \neq n$. Another necessary condition for thermalization is that the fluctuations with respect to $\bar{O}$ must be small in the TDL. The prevailing paradigm for the thermalization mechanism is the Eigenstate Thermalization Hypothesis \cite{Rigol08}, but more general ideas have also been proposed \cite{Sirker14}.
To calculate the canonical thermal ensemble $\rho_{th}$ that describes the equilibrated long time limit the energy $\langle \Psi(0) \vert H \vert \Psi(0) \rangle$ is set equal to the thermal average $Tr(H \rho_{th})$, and the equation is numerically solved for the temperature.


\section{Scaling and Time Dependence of Local Correlations}
\label{subsec:scalingcorrelations}

\begin{figure}
\includegraphics[width=4in,height=3in]{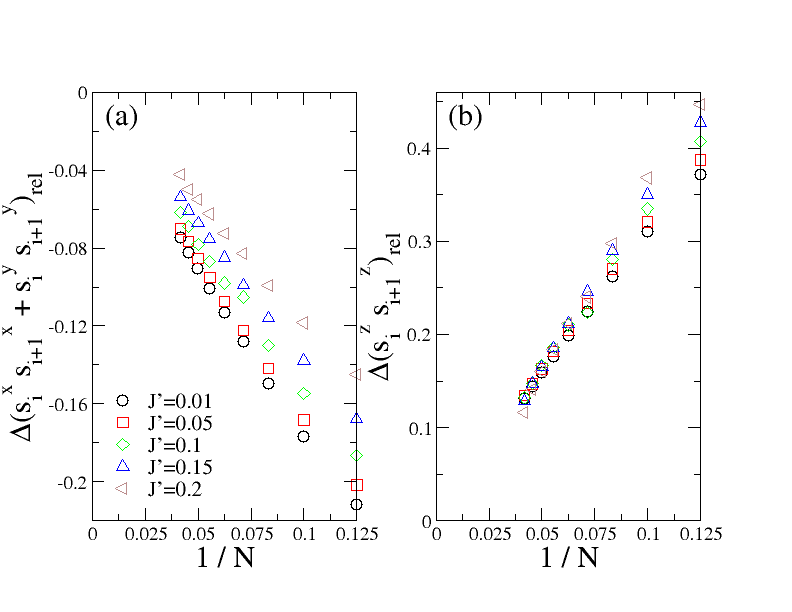}
\vspace{0pt}
\caption{Scaling of the difference $\Delta\hat{O}_{rel}=\frac{\bar{O} - \langle \hat{O} \rangle _{th}}{\langle \hat{O} \rangle _{th}}$ between the diagonal and canonical thermal ensemble values divided by the canonical thermal ensemble value as function of inverse length $\frac{1}{N}$ for correlation function (a) $\hat{O} = s_i^x s_{i+1}^x + s_i^y s_{i+1}^y$ and (b) $\hat{O} = s_i^z s_{i+1}^z$ for $\Delta=1.1$ and different values of $J'$. The initial state is the ground state for $\Delta_0=15$. The maximum chain has $N=24$.
}
\label{fig:Neelscalingsisi+1both}
\end{figure}

\begin{figure}
\includegraphics[width=4in,height=3in]{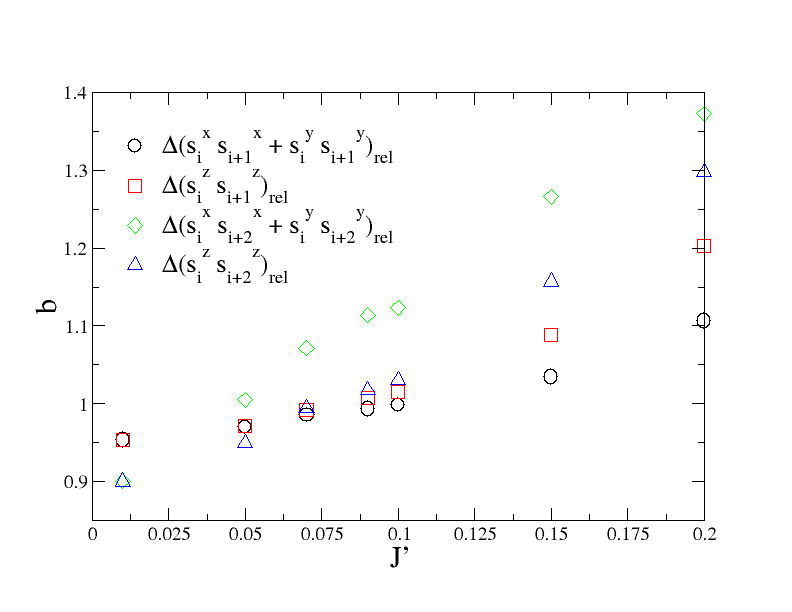}
\vspace{0pt}
\caption{Exponent $b$ of a power law $\Delta\hat{O}_{rel}=a(\frac{1}{N})^b$ fitted through $\Delta\hat{O}_{rel}=\frac{\bar{O} - \langle \hat{O} \rangle _{th}}{\langle \hat{O} \rangle _{th}}$ as function of $J'$ with $\Delta=1.1$ (see Figs \ref{fig:Neelscalingsisi+1both} and \ref{fig:Neelscalingsisi+2both}). The initial state is the ground state for $\Delta_0=15$.
}
\label{fig:exponentialxmgrace}
\end{figure}

To investigate how the rate of thermalization depends on the integrability breaking parameter $J'$ of Hamiltonian (\ref{eq:Hchain}) the difference of the diagonal and canonical thermal ensemble values for a local operator $\hat{O}$ is calculated as function of the size of the chain $N$. More specifically, the relative difference of the two ensembles is defined as the difference in their values divided by the value of the canonical thermal ensemble, so that $\Delta\hat{O}_{rel}=\frac{\bar{O} - \langle \hat{O} \rangle _{th}}{\langle \hat{O} \rangle _{th}}$.
The scaling of $\Delta(s_i^x s_{i+1}^x + s_i^y s_{i+1}^y)_{rel}$ with $\frac{1}{N}$ is shown in Fig. \ref{fig:Neelscalingsisi+1both}(a) as function of $J'$ for $N \leq 24$. $\Delta(s_i^x s_{i+1}^x + s_i^y s_{i+1}^y)_{rel}$ is getting smaller in magnitude with increasing $N$ and appears to be directed towards the origin in the TDL where $\frac{1}{N} \to 0$, which agrees with the expectation that the Hamiltonian of Eq. (\ref{eq:Hchain}) thermalizes as long as $J' \neq 0$. If a power law dependence $a(\frac{1}{N})^b$ is fitted through the data for each $J'$ value, Fig. \ref{fig:exponentialxmgrace} shows the dependence of the exponent $b$ on $J'$.
For very small $J'$ it is $b<1$ indicating that convergence to the thermal ensemble is slow for weak integrability breaking. As $J'$ increases $b$ becomes approximately 1 around $J' \sim 0.1$, showing that convergence becomes linear and thermalization faster. For even stronger $J'$ it is $b>1$ and thermalization becomes even faster.

\begin{figure}
\includegraphics[width=4in,height=3in]{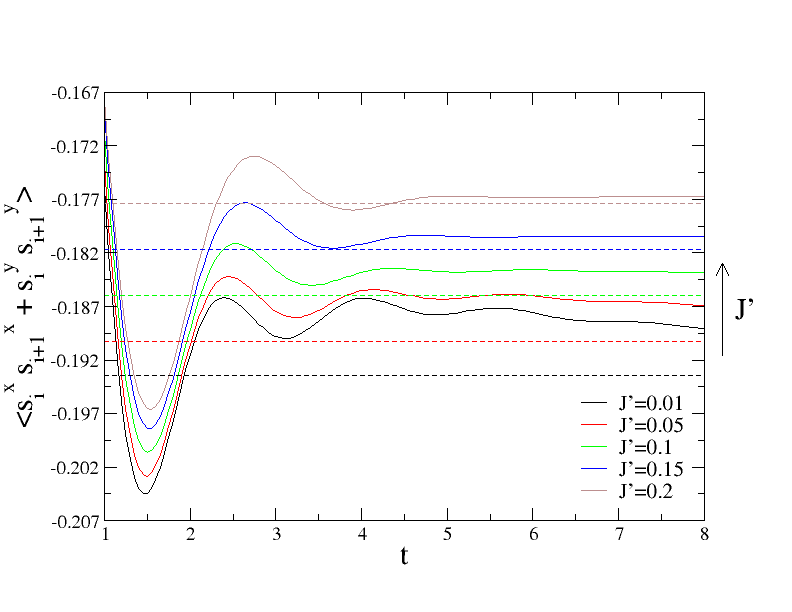}
\vspace{0pt}
\caption{Time evolution of $<s_i^x s_{i+1}^x + s_i^y s_{i+1}^y>$ for different values of $J'$ with $\Delta=1.1$. The dashed lines correspond to the values of the canonical thermal ensemble $<s_i^x s_{i+1}^x + s_i^y s_{i+1}^y>_{\textrm{th}}$. The initial state is the ground state for $\Delta_0=15$.
}
\label{fig:Neeltimesisi+1}
\end{figure}
\begin{figure}
\includegraphics[width=4in,height=3in]{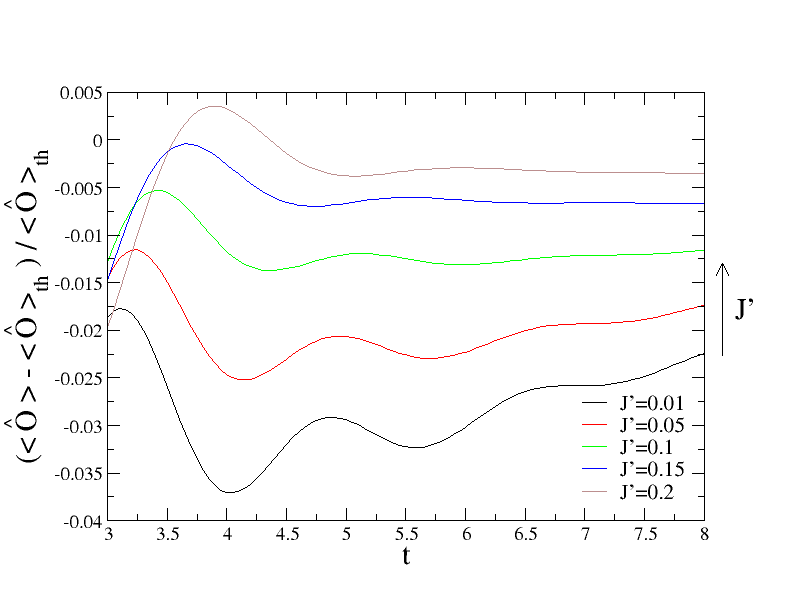}
\vspace{0pt}
\caption{Time evolution of the expectation value of $\hat{O} = s_i^x s_{i+1}^x + s_i^y s_{i+1}^y$ with respect to its thermal value $<\hat{O}>_{th}$ for different values of $J'$ with $\Delta=1.1$. The initial state is the ground state for $\Delta_0=15$.
}
\label{fig:Neeltimesisi+1diff}
\end{figure}

These results can be corroborated by direct calculation of the time evolution of $<s_i^x s_{i+1}^x + s_i^y s_{i+1}^y>$ with the Chebyshev polynomial expansion \cite{Zhang07,Zhang08}, which can treat larger chains in comparison with exact diagonalization. This data is shown in Fig. \ref{fig:Neeltimesisi+1}, and the difference between the time evolved $<s_i^x s_{i+1}^x + s_i^y s_{i+1}^y>$ and the canonical thermal ensemble value scaled with the canonical thermal ensemble value is shown in Fig. \ref{fig:Neeltimesisi+1diff}. The finite size of the system eventually generates revivals of the operator expectation value in later times, and the closed chain with $N=30$ used here generates the infinite chain results for times at least up to $\frac{8}{J}$ \cite{Essler13}. This has also been vindicated by comparison of the results as function of $N$, which agree to progressively longer times with increasing $N$. The slow thermalization close to the integrable limit has brought forward the concept of prethermalization, where local operator expectation values go through a prethermalized regime where they first relax to a non-thermal quasi-stationary value before reaching the canonical thermal ensemble value \cite{Kollar11,Essler13}. This does not imply absence of thermalization, but rather that fluctuations are more slowly varying in time in comparison with stronger integrability breaking. Here it is also observed that for weak $J' \lesssim 0.05$ (Figs \ref{fig:Neeltimesisi+1} and \ref{fig:Neeltimesisi+1diff}) there is slow convergence of the time expectation value to the thermal ensemble, which agrees with the results for the scaling of $\Delta(s_i^x s_{i+1}^x + s_i^y s_{i+1}^y)_{rel}$ with $\frac{1}{N}$ that gave an exponent $b<1$ in the power law fit. For stronger $J'=0.1$ where the finite size scaling of $\Delta(s_i^x s_{i+1}^x + s_i^y s_{i+1}^y)_{rel}$ points to faster thermalization Figs \ref{fig:Neeltimesisi+1} and \ref{fig:Neeltimesisi+1diff} also show that,
with the time dependent difference tending to zero more quickly and the smaller time fluctuations becoming weaker. For the highest values of $J'$ Figs \ref{fig:Neeltimesisi+1} and \ref{fig:Neeltimesisi+1diff} show an even higher thermalization rate, in agreement with the scaling data. The difference between the time evolved $<s_i^x s_{i+1}^x + s_i^y s_{i+1}^y>$ and the thermal ensemble is becoming smaller with an even faster rate.

\begin{figure}
\includegraphics[width=4in,height=3in]{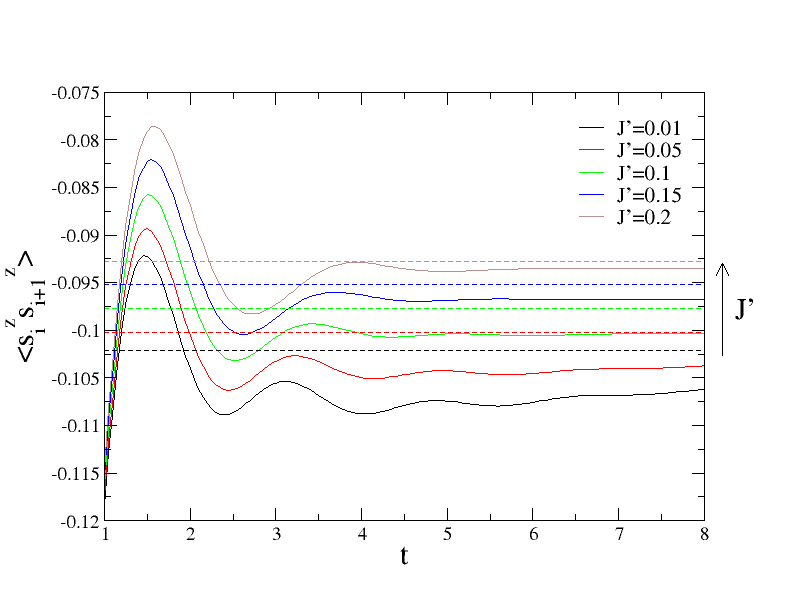}
\vspace{0pt}
\caption{Time evolution of $<s_i^z s_{i+1}^z>$ for different values of $J'$ with $\Delta=1.1$. The dashed lines correspond to the values of the canonical thermal ensemble $<s_i^z s_{i+1}^z>_{\textrm{th}}$. The initial state is the ground state for $\Delta_0=15$.
}
\label{fig:Neeltimesziszi+1}
\end{figure}

A similar procedure for the nearest neighbor correlation function $s_i^z s_{i+1}^z$ generates the scaling data of Fig. \ref{fig:Neelscalingsisi+1both}(b).
Again $\Delta(s_i^z s_{i+1}^z)_{rel}$ decreases with $N$ and appears to be directed towards the origin in the TDL where $\frac{1}{N} \to 0$. A power law fit $a(\frac{1}{N})^b$ shows that the rate of thermalization changes as function of integrability breaking in exactly the same way as with correlation $s_i^x s_{i+1}^x + s_i^y s_{i+1}^y$, with the exponent $b$ having the same dependence on $J'$ and being close in value to the one of $\Delta(s_i^x s_{i+1}^x + s_i^y s_{i+1}^y)_{rel}$ (Fig. \ref{fig:exponentialxmgrace}). The time-dependent data of Fig. \ref{fig:Neeltimesziszi+1} also show that for the maximum available time the difference between the expectation value $<s_i^z s_{i+1}^z>$ and the canonical thermal ensemble value decreases with $J'$.

\begin{figure}
\includegraphics[width=4in,height=3in]{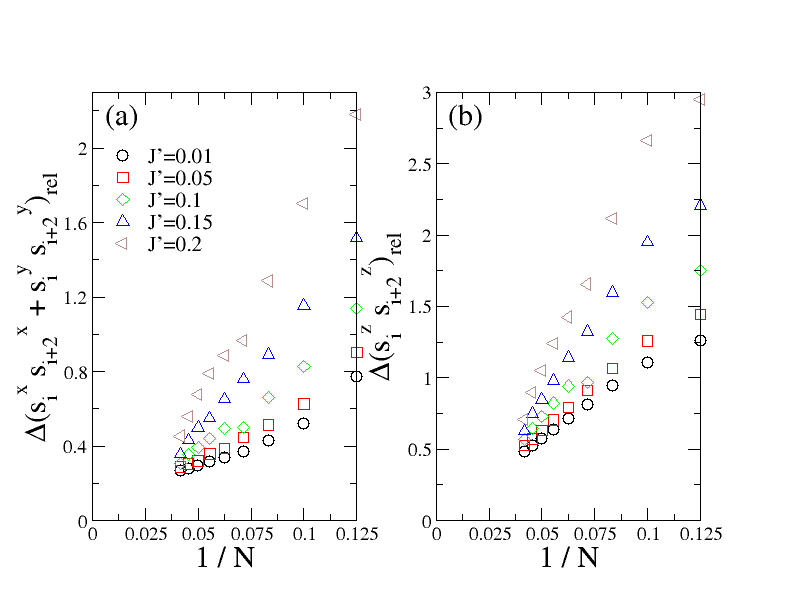}
\vspace{0pt}
\caption{Scaling of the difference $\Delta\hat{O}_{rel}=\frac{\bar{O} - \langle \hat{O} \rangle _{th}}{\langle \hat{O} \rangle _{th}}$ between the diagonal and canonical thermal ensemble values divided by the canonical thermal ensemble value as function of inverse length $\frac{1}{N}$ for correlation function (a) $\hat{O}=s_i^x s_{i+2}^x + s_i^y s_{i+2}^y$ and (b) $\hat{O}=s_i^z s_{i+2}^z$ for $\Delta=1.1$ and different values of $J'$. The initial state is the ground state for $\Delta_0=15$. The maximum chain has $N=24$.
}
\label{fig:Neelscalingsisi+2both}
\end{figure}

\begin{figure}
\includegraphics[width=4in,height=3in]{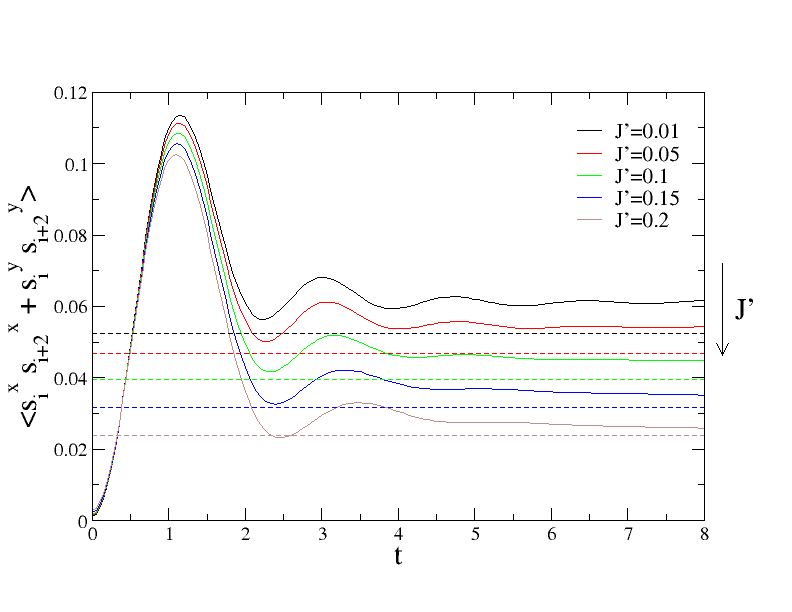}
\vspace{0pt}
\caption{Time evolution of $<s_i^x s_{i+2}^x + s_i^y s_{i+2}^y>$ for different values of $J'$ with $\Delta=1.1$. The dashed lines correspond to the values of the canonical thermal ensemble $<s_i^x s_{i+2}^x + s_i^y s_{i+2}^y>_{\textrm{th}}$. The initial state is the ground state for $\Delta_0=15$.
}
\label{fig:Neeltimesisi+2}
\end{figure}

\begin{figure}
\includegraphics[width=4in,height=3in]{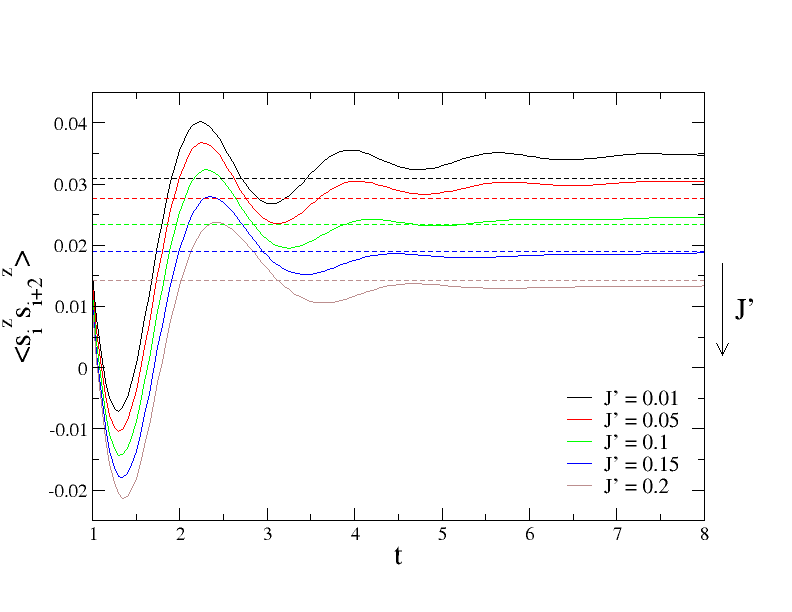}
\vspace{0pt}
\caption{Time evolution of $<s_i^z s_{i+2}^z>$ for different values of $J'$ with $\Delta=1.1$. The dashed lines correspond to the values of the canonical thermal ensemble $<s_i^z s_{i+2}^z>_{\textrm{th}}$. The initial state is the ground state for $\Delta_0=15$.
}
\label{fig:Neeltimesziszi+2}
\end{figure}

Considering next-nearest neighbor correlations, the dependence of $\Delta(s_i^x s_{i+2}^x + s_i^y s_{i+2}^y)_{rel}$ and $\Delta(s_i^z s_{i+2}^z)_{rel}$ on $\frac{1}{N}$ is shown in Fig. \ref{fig:Neelscalingsisi+2both}. They decrease in magnitude as $N$ increases, and the exponent $b$ of a fit to a function of the form $a(\frac{1}{N})^b$ is plotted in Fig. \ref{fig:exponentialxmgrace} as function of $J'$. Again three different regimes with respect to the rate of thermalization as function of $N$ are seen, with the exponents for the two operators not differing much. The time-dependent data of Figs \ref{fig:Neeltimesisi+2} and \ref{fig:Neeltimesziszi+2} confirm the similarity with the nearest neighbor operator case, with the difference between the time-dependent expectation values and the canonical thermal ensemble value becoming smaller when going away from the integrable limit as thermalization becomes faster.

Comparing the discrepancy of the time expectation values from the canonical thermal emsemble value, it is seen from Figs \ref{fig:Neeltimesisi+1} and \ref{fig:Neeltimesisi+2} and Figs \ref{fig:Neeltimesziszi+1} and \ref{fig:Neeltimesziszi+2} that the nearest neighbor correlations approach their thermalized limit faster than the next-nearest neighbor correlations for the longest times available. This is because nearest neighbor correlations are more local, therefore the rest of the system acts as a bigger and more effective bath, at least for the sizes available for the calculations.

Comparing the value of the fitting exponent $b$ between the two cases in Fig. \ref{fig:exponentialxmgrace}, for weak $J'$ the exponent is smaller for next-nearest neighbor correlations. This has to do again with the more local nature of nearest neighbor correlations. Since convergence to the TDL limit is slow, it will take a larger $N$ before the effective bath size becomes equivalent for correlations betweens spins one and two sites way. With increasing $J'$ the exponent $b$ becomes higher for next-nearest neighbor correlations. For higher $J'$ the interactions are becoming less local and $J'$ directly connects spins two sites apart in a more efficient way. Thus progressively a smaller system size is required for the effective bath size to be the same for two and three sites, consequently $b$ is higher for next-nearest neighbors as they approach this system size fast on the way to the TDL.

\section{Off-Diagonal Terms and Thermalization}
\label{subsec:offdiagonalterms}

As discussed in Sec. \ref{sec:timeevolution}, Eq. (\ref{eq:diagonal}) demonstrates that the off-diagonal terms $\hat{O}_{nm}$ determine the rate of equilibration and the strength of fluctuations around the infinite time limit. Quantum chaos theory predicts that the $\hat{O}_{nm}$ are characterized by a smooth distribution \cite{Srednicki96,Srednicki99}. More precisely, the fluctuations in Eq. (\ref{eq:diagonal}) are determined by the strength of the $\hat{O}_{nm}$ and the phases $E_n-E_m \equiv \Delta E_{nm}$. High frequency fluctuations involve eigenstates with large $\Delta E_{nm}$. At longer times these terms average out to zero and the long time limit is determined by overlaps between eigenstates close in energy. The relative strength of $\hat{O}_{nm}$'s with small and larger energy differences is therefore crucial for the rate of equilibration. The distribution of the $(s_i^x s_{i+1}^x + s_i^y s_{i+1}^y)_{nm}$ with $\Delta E_{nm}$ depends on the subsector of the Hamiltonian that time evolution takes place and if all symmetries have been taken into account in the time evolution, which is determined by the integrability or not of Hamiltonian (\ref{eq:Hchain}), and also on the specific values of the Hamiltonian parameters, which determine the strength of the overlaps between different eigenstates.

The Hamiltonian subsector where time evolution takes place is determined by the ground state before the quench. This subsector belongs to a specific one-dimensional irreducible representation of the total symmetry group of Hamiltonian (\ref{eq:Hchain}). However, in the integrable limit there is a total number of $N$ conserved quantities which are not fully taken into account. Consequently the Hamiltonian subsector that determines time evolution has states that would belong to different subsectors had all the conservation laws been taken into account. The nearest neighbor correlation functions $s_i^x s_{i+1}^x + s_i^y s_{i+1}^y$ and $s_i^z s_{i+1}^z$ are translationally invariant in singly degenerate subsectors and qualitatively similar to the nearest neighbor energy (it is noted that their sum does not correspond to the energy when $\Delta \neq 1$). A Hamiltonian subsector where all symmetries have been taken into account is characterized by level repulsion and has an energy difference between any pair of states which is in general not close to zero, provided there are no small parameters in the Hamiltonian. As in the integrable limit many of the conservation laws are not taken into account in the block diagonalization of the Hamiltonian it is expected that there will be levels very close in energy, which would belong to different subsectors of the Hamiltonian if the extra conservation laws were to be taken into account. Because of this reason the Hamiltonian matrix elements between such levels should be close to zero, and consequently the same should hold for the off-diagonal matrix elements of correlations $s_i^x s_{i+1}^x + s_i^y s_{i+1}^y$ and $s_i^z s_{i+1}^z$. For this reason in the integrable limit the distribution of the off-diagonal elements of the spin correlation functions has small values close to zero and peaks at an energy difference away from zero.

\begin{figure}
\includegraphics[width=4in,height=3in]{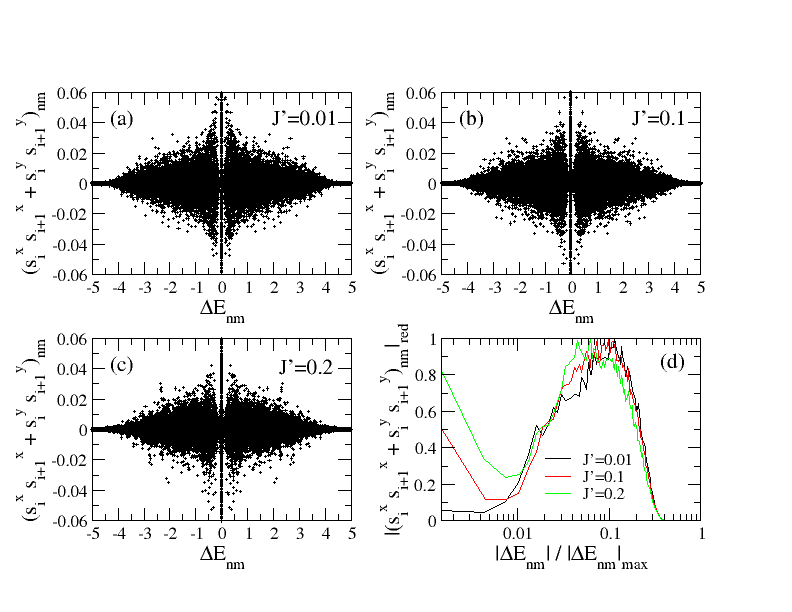}
\vspace{0pt}
\caption{Distribution of matrix elements $(s_i^x s_{i+1}^x + s_i^y s_{i+1}^y)_{nm}$ as function of the energy difference $\Delta E_{nm} = E_n-E_m$ for (a) $J'=0.01$, (b) $J'=0.1$, and (c) $J'=0.2$. The distribution is symmetric with respect to the $\Delta E_{nm}$ axis. The coarse grained average of $|(s_i^x s_{i+1}^x + s_i^y s_{i+1}^y)_{nm}|$ divided with its maximum value is plotted in (d) as function of the reduced average energy difference for 100 bins with an equal number of points.
}
\label{fig:Neelcorrfncsisi+1}
\end{figure}

Once integrability is broken with $J' \neq 0$ the additional conservation laws of the integrable case do not apply anymore. States with small energy differences start to mix more strongly and the correlation functions start to develop significant off-diagonal matrix elements between such states, especially since according to low order perturbation theory the mixing is stronger for levels having close energies in the integrable limit. The strength of these off-diagonal elements is regulated by the magnitude of the integrability breaking term. A plot of the $(s_i^x s_{i+1}^x + s_i^y s_{i+1}^y)_{nm}$ as function of $\Delta E_{nm}$ is shown in Fig. \ref{fig:Neelcorrfncsisi+1} for $N=18$ for $J'=0.01$, $0.1$ and $0.2$, along with a coarse grained average of the matrix elements as function of $\Delta E_{nm}$ in Fig. \ref{fig:Neelcorrfncsisi+1} (d).
For weak $J'=0.01$ the corresponding distribution of Fig. \ref{fig:Neelcorrfncsisi+1}(a) and the coarse grained average in Fig. \ref{fig:Neelcorrfncsisi+1}(d) peak away from the origin as function of $\Delta E_{nm}$, with a much weaker low energy peak induced by the integrability breaking term. According to Eq. (\ref{eq:diagonal}) since the high-frequency fluctuations have a significantly higher strength it takes a considerable time for them to sufficiently diminish due to the mixing of their different phases and only then the significantly weaker low frequency modes dominate time evolution. For the short times available to numerical calculations it is not possible to see later times where the long time limit has settled in, but only a transient regime where the system has not yet fully thermalized and the expectation value has not fully converged to its infinite time limit (Figs \ref{fig:Neeltimesisi+1} and \ref{fig:Neeltimesisi+1diff}). A bigger $J'=0.1$ generates stronger $(s_i^x s_{i+1}^x + s_i^y s_{i+1}^y)_{nm}$'s for low energy differences and for even higher $\Delta E_{nm}$ as evidenced by Figs \ref{fig:Neelcorrfncsisi+1}(b) and \ref{fig:Neelcorrfncsisi+1}(d). The strength of the low energy peak becomes of the same order of magnitude with the high energy peak and thermalization is now faster.
The effects become even stronger for $J'=0.2$. The low-energy peak acquires strength comparable to the higher energy peak and the two peaks also begin to merge (Figs \ref{fig:Neelcorrfncsisi+1}(c) and (d)). This results in the qualitative change in the time evolution of $<s_i^x s_{i+1}^x + s_i^y s_{i+1}^y>$ and its faster thermalization. The above provide the explanation for the scaling of the correlations functions and the time-dependent data.
It must be noted that the above considerations do not take into account the energy distribution of the coefficients of the initial state $C_n$ (Eq. (\ref{eq:diagonal})), however their distribution has bigger weight for smaller $\Delta E_{nm}$ as they approach a sharp peak going to the TDL, therefore at most renormalize
the low energy peak of the $(s_i^x s_{i+1}^x + s_i^y s_{i+1}^y)_{nm}$ distribution and strengthen it with respect to the higher energy peak.



\section{Conclusions}
\label{sec:conclusions}
The approach to thermalization as function of integrability breaking was investigated for the one-dimensional AAHM with nearest and next-nearest neighbor interactions, with spatial and spin symmetries taken into account. The scaling of the difference of the diagonal and canonical thermal ensembles was examined for local operators, whose time evolution was also calculated for even larger systems. Both pointed to a rate of thermalization depending on the strength of the next-nearest neighbor interactions $J'$. Three different regimes were identified, the first for weak $J'$ where thermalization is slow and the dependence of the difference of the two ensembles on $\frac{1}{N}$ is sublinear, then a second for stronger $J'$ where thermalization is faster and the difference is approximately linear in $\frac{1}{N}$, and then a third where thermalization is even faster for even larger $J'$ with superlinear dependence of the difference on $\frac{1}{N}$. The existence of three different thermalization regimes was explained by the dependence of the off-diagonal elements of the operators in question on the energy difference of the corresponding eigenstates. They were shown to develop a low energy peak that determines the long time fluctuations, grows stronger with $J'$ and competes with the high energy peak facilitating faster thermalization as its strength increases.

The author is grateful to A. Lazarides for many suggestions and comments, and thanks D. Schuricht for a discussion.

\bibliography{paperthermalization}

\end{document}